\documentclass[sigconf]{acmart}

\usepackage{amsmath}
\usepackage{amsthm}
\usepackage{balance}

\usepackage{graphicx}
\usepackage{multirow}
\usepackage{xspace}
\usepackage{subfigure}
\usepackage{amsmath}
\usepackage{enumitem}
\usepackage[linesnumbered,ruled]{algorithm2e}
\usepackage{wrapfig}
\usepackage{ulem}
\newtheorem{theorem}{Theorem}
\theoremstyle{definition}
\newtheorem{definition}{Definition}[section]
\newcommand{\ourmethod}{FreqD\xspace}

\definecolor{bgc}{HTML}{DBE7FA}

\AtBeginDocument{%
  \providecommand\BibTeX{{%
    \normalfont B\kern-0.5em{\scshape i\kern-0.25em b}\kern-0.8em\TeX}}}

\copyrightyear{2025}
\acmYear{2025}
\setcopyright{acmlicensed}\acmConference[KDD '25]{Proceedings of the 31st ACM SIGKDD Conference on Knowledge Discovery and Data Mining V.1}{August 3--7, 2025}{Toronto, ON, Canada}
\acmBooktitle{Proceedings of the 31st ACM SIGKDD Conference on Knowledge Discovery and Data Mining V.1 (KDD '25), August 3--7, 2025, Toronto, ON, Canada}
\acmDOI{10.1145/3690624.3709248}
\acmISBN{979-8-4007-1245-6/25/08}

\begin{document}


\title{Exploring Feature-based Knowledge Distillation for Recommender System: A Frequency Perspective}


\author{Zhangchi Zhu}
\affiliation{%
  \institution{East China Normal University}
  \city{Shanghai}
  \country{China}
}
\email{zczhu@stu.ecnu.edu.cn}

\author{Wei Zhang}
\authornote{Corresponding author.}
\affiliation{%
  \institution{East China Normal University}
  \city{Shanghai}
  \country{China}
}
\email{zhangwei.thu2011@gmail.com}

\begin{abstract}
  In this paper, we analyze the feature-based knowledge distillation for recommendation from the frequency perspective. By defining knowledge as different frequency components of the features, we theoretically demonstrate that regular feature-based knowledge distillation is equivalent to equally minimizing losses on all knowledge and further analyze how this equal loss weight allocation method leads to important knowledge being overlooked. In light of this, we propose to emphasize important knowledge by redistributing knowledge weights. Furthermore, we propose \ourmethod, a lightweight knowledge reweighting method, to avoid the computational cost of calculating losses on each knowledge. Extensive experiments demonstrate that \ourmethod consistently and significantly outperforms state-of-the-art knowledge distillation methods for recommender systems. Our code is available at \url{https://github.com/woriazzc/KDs}.
\end{abstract}




\maketitle

\section{Introduction}\label{sec:intro}

With the rapid growth in the variety of items and the gradual diversification of user preferences, recent studies~\cite{ohsaka2023curse,liang2018variational} have begun to increase the model capacity by increasing the dimension of user and item embeddings. However, due to the huge share of embeddings~\cite{ohsaka2023curse,zhao2023embedding} in the overall model parameters, increasing the embedding dimensions greatly increases the number of parameters, which in turn increases the storage cost and inference latency, leading to longer waiting times and lower user satisfaction.

To improve the inference efficiency without sacrificing accuracy, many studies~\cite{kang2020rrd,ShiLWZ23,kang2021topology,kang2022personalized} have adopted \textit{Knowledge Distillation} (KD) to recommender system.
KD is a model-agnostic approach for model compression~\cite{hinton2015distilling,gou2021knowledge}.
In knowledge distillation for recommendation, the common process is first to train a large teacher model using the user-item interactions, then train a small student model using the user-item interactions as well as the features in the intermediate layer~\cite{kang2020rrd,kang2021topology,kang2022personalized} and the predictions in the output layer~\cite{lee2019collaborative,kang2020rrd,kweon2021bidirectional,chen2023unbiased} provided by the teacher model. After training, the student trained with KD performs similarly to the teacher and has a much lower inference latency due to its small representation dimensionality. Most KD methods for recommendation~\cite{kweon2021bidirectional,tang2018ranking,chen2023unbiased,lee2019collaborative} are response-based methods that force student models to learn the teachers' logits to improve their performance. Recently, some feature-based KD methods~\cite{kang2020rrd,kang2022personalized} are proposed. They use the teacher's hidden representations as additional supervisory signals and have achieved promising performance. However, unlike the observations in CV and NLP, these feature-based methods behave much worse than response-based methods, as shown in Table~\ref{tab:all}. This is very counter-intuitive, as the mainstream studies in this regard~\cite{deng2019deepcf,he2020lightgcn,wang2022towards} show that features play a crucial role in recommender systems. 

We believe the above phenomenon is due to the lack of a fundamental understanding of the knowledge transfer processes during feature-based distillation. To this end, our work focuses on the feature-based knowledge distillation methods for recommender systems and aims to investigate the following research questions:
\begin{itemize}[leftmargin=*]
    \item[1)] \textbf{What} ``knowledge'' in features is transferred from teachers to student recommendation models when minimizing feature-based knowledge distillation loss?
    \item[2)] Considering the huge gap between the capacity of the student and teacher, \textbf{why} is the important knowledge in features not being successfully transferred to the student recommendation models?
    \item[3)] \textbf{How} to transfer as much important knowledge in features as possible to student recommendation models while avoiding increasing the complexity of knowledge distillation?
\end{itemize}

To answer the \textbf{first question}, we first need to define the knowledge contained in the features. Therefore, we need a rigorous system for analyzing the features. Thanks to the development of graph signal processing~\cite{kipf2016semi,wu2023extracting} and graph-based collaborative filtering~\cite{he2020lightgcn,peng2022less,choi2023blurring}, we can get a clearer picture of the features from the frequency domain perspective. Specifically, recent studies~\cite{choi2023blurring,wu2023extracting,peng2022less} have demonstrated that the low-frequency component of a feature characterizes the common properties between a node (user or item) and its neighboring nodes, whereas the high-frequency component characterizes the differences between nodes. We, therefore, define various types of knowledge as different frequency components of features. Then, we theoretically demonstrate that we equally transfer all types of knowledge from the teacher recommendation models to student recommendation models when minimizing feature-based knowledge distillation loss.

To answer the \textbf{second question}, we review the training process of the feature-based knowledge distillation and investigate the results of student models learning various types of knowledge. We also investigate the mastery of knowledge by student models of different capacities. From the experimental results shown in Figure~\ref{fig:fd_loss_diffs}, we find that: (1) the low-frequency components of features (which characterize the commonalities between nodes and their neighbors) are more difficult to master by student models compared to the high-frequency components; (2) the mastery of the low-frequency components gets worse as the capacity of student models decreases. In addition, we show that different knowledge plays different roles by removing each type of knowledge separately from the feature-based distillation loss. As can be seen from the recommendation performance shown in Table~\ref{tab:wo_diffk}, low-frequency components play a more critical role than high-frequency components. These results demonstrate that the low-frequency knowledge that is important to students is not well transferred.

For the \textbf{third question}, based on our analysis in the second question, we propose to redistribute the weights for distilling the various types of knowledge. We design three schemes for weight assignment: (1) constant weights, (2) weights positively correlated with frequency, and (3) weights negatively correlated with frequency. We empirically show that by setting weights negatively correlated with frequency, we can effectively improve the mastery of student models for the low-frequency components. However, it requires a complete eigendecomposition of the graph Laplace matrix, which is impractical in a real scenario due to the huge number of users and items. To avoid increasing the complexity of the feature-based approach, we further propose \ourmethod, a lightweight knowledge reweighting approach, which transforms the above plain practice into graph filtering of teacher and student features separately before knowledge distillation. We also theoretically demonstrate the equivalence of the two methods.

To summarize, this work makes the following contributions:
\begin{itemize}[leftmargin=*]
    \item By interpreting the knowledge in features from a frequency perspective, we reveal for the first time the underlying mechanism of feature-based knowledge distillation for recommender systems.
    \item We empirically investigate: (1) the mastery of different types of knowledge by student models with different capacities, and (2) the distinct importance of various types of knowledge. Based on this, we emphasize the significance of improving the student recommendation models' mastery of knowledge revealed in low-frequency components.
    \item We propose a new weight assignment scheme to help student models better grasp the knowledge in low-frequency components. Further, we propose \ourmethod, a lightweight knowledge reweighting method, to achieve weight reassignment without increasing the complexity of feature-based knowledge distillation.
    \item We conduct extensive experiments on three public datasets and use three backbones to demonstrate the superiority of our \ourmethod in improving recommendation performance.
\end{itemize}

\section{Related Work}

\subsection{Graph-based Collaborative Filtering}

With the increasing development of graph neural networks~\cite{wu2020comprehensive,xu2018powerful,scarselli2008graph,zhang2019heterogeneous}, researchers have introduced them into the field of recommender systems~\cite{wu2022graph,ricci2021recommender,wang2021survey}. Graph-based collaborative filtering (GCF) aims to conduct collaborative filtering with graph processing. It provides a general framework to incorporate high-order information in the user-item graph to boost the recommendation performance. NGCF~\cite{wang2019neural} first proposes the graph collaborative filtering framework to make recommendations. LightGCN~\cite{he2020lightgcn} empirically shows the uselessness of non-linearity and feature transformation. Recently, some works have analyzed and improved graph models from a spectral perspective. For example, GF-CF~\cite{shen2021powerful} theoretically shows that LightGCN and some traditional CF methods are essentially low-pass filters. FF-G2M~\cite{wu2023extracting} factorizes features into low- and high-frequency components, demonstrating they represent commonalities and differences between nodes, respectively. GDE~\cite{peng2022less} proposes that low-frequency components are much more important than high-frequency components in recommender systems.

\subsection{Knowledge Distillation}

Knowledge distillation (KD) is a model compression method that distills the knowledge from a well-trained teacher model to a small student model. Existing KD methods mainly fall into three categories~\cite{gou2021knowledge} based on their knowledge type: response-based methods, feature-based methods, and relation-based methods. \textit{Response-based} methods~\cite{hinton2015distilling,zhao2022decoupled,huang2022knowledge} take the final prediction of the teacher as soft targets, and the student is trained to mimic the soft targets by minimizing the divergence loss. \textit{Feature-based} methods take advantage of the teacher model's ability to learn multiple levels of feature representation with increasing abstraction. They propose to enrich auxiliary signals by matching the learned features of the student and the teacher. These methods leverage various types of knowledge, such as feature activations~\cite{romero2014fitnets}, attention maps~\cite{zagoruyko2016paying}, probability distribution in feature space~\cite{passalis2018learning}. \textit{Relation-based} methods~\cite{park2019relational,chen2020learning,passalis2020probabilistic} further explore the relationships between different layers or data samples. A representative example is RKD~\cite{park2019relational}, which calculates the similarity between each pair of samples to obtain a similarity matrix and then minimizes the difference between the teacher's and the student's similarity matrices.

\subsection{Knowledge Distillation in Recommender System}

Knowledge distillation has been introduced into recommender systems to reduce the representation dimensionality and inference latency, which falls into the same three categories, i.e., response-based, feature-based, and relation-based.

\textit{Response-based} methods focus on teachers' predictions. For example, CD~\cite{lee2019collaborative} samples unobserved items from a distribution associated with their rankings predicted by students: items with higher rankings are more likely to be sampled. RRD~\cite{kang2020rrd} adopts a list-wise loss to maximize the likelihood of teachers' recommendation list. DCD~\cite{lee2021dual} samples items and users from distributions associated with teacher-student prediction discrepancy to provide the student with dynamic knowledge. UnKD~\cite{chen2023unbiased} partitions items into multiple groups based on popularity and samples positive-negative item pairs within each group. HetComp~\cite{kang2023distillation} aims to transfer the ensemble knowledge of heterogeneous teachers. It guides the student model by transferring easy-to-hard knowledge sequences generated from the teachers’ trajectories. CCD~\cite{lee2024continual} focuses on KD in a non-stationary data stream, where both the teacher and the student continually and collaboratively evolve along the data stream. It makes the student and teacher promote each other.

\textit{Feature-based} methods focus on the intermediate representations of the teacher. For example, DE~\cite{kang2020rrd} proposes an expert module comprising $K$ projectors and a selection network. The selection network assigns each item or user to one of the projectors according to the teacher's representation of each. Then, the MSE loss is leveraged to align the teacher representation with the student representation transformed by the selected projector. PHR~\cite{kang2022personalized} employs a personalization network that enables a personalized distillation for each user/item representation, which can be viewed as a generalization of DE. However, existing feature-based methods ignore differences in the different frequencies of the features.

\textit{Relation-based} methods focus on the relationships between different entities (i.e., users and items). HTD~\cite{kang2021topology} observes that the student is too small to learn the whole relationship. As a result, the vanilla relation-based distillation approach is not always effective and even degrades the student’s performance. Therefore, it proposes to distill the sample relation hierarchically to alleviate the large capacity gap between the student and the teacher.
Since relation-based knowledge distillation
methods are also closely related to features, we discuss how the main idea of this paper could be applied to them in Section~\ref{sec:discuss}. 


\section{Preliminary}

\subsection{Top-N Recommendation}
In this work, we focus on the top-$N$ recommendation with implicit feedback. 
Specifically, let $\mathcal{U}$ and $\mathcal{I}$ denote the user set and item set, respectively. Then $|\mathcal{U}|$ and $|\mathcal{I}|$ are taken as the number of users and items, respectively. The historical implicit feedback can be formulated as a set of observed user-item interactions $\mathcal{R}=\{(u, i)|u$ interacted with $i\}$. A recommendation model aims to score the items not interacted with by the user and recommend $N$ items with the largest scores. 
As for the learning objective, most works adopt the BPR loss~\cite{rendle2012bpr} to make the predicted scores of interacted items higher than randomly sampled negative items.

\subsection{Graph Signal Processing}
In this paper, we use tools in graph signal processing to define the knowledge in features. Given a graph $\mathcal{G}=(\mathcal{V},\mathcal{E})$ where $\mathcal{V}$ and $\mathcal{E}$ being the node set and edge set, respectively. We denote $|\mathcal{V}|$ and $|\mathcal{E}|$ the number of nodes and the number of edges, respectively. Graph signal processing performs the graph Fourier transformation on a graph signal to decompose it onto different frequency bands. Specifically, let $\mathbf{A}\in \mathbb{R}^{|\mathcal{V}|\times |\mathcal{V}|}$ be the adjacency matrix of $\mathcal{G}$ and $\mathbf{D}=\text{Diag}(\mathbf{A}\mathbf{1})$ be the degree matrix with $\mathbf{1}\in\mathbb{R}^{|\mathcal{V}|}$ being the all-one vector. The symmetric normalized graph Laplacian matrix is given by $\tilde{\mathbf{L}}=\mathbf{I}-\mathbf{D}^{-1/2}\mathbf{A}\mathbf{D}^{-1/2}$.
As $\tilde{\mathbf{L}}$ is real and symmetric, its eigendecomposition is given by $\tilde{\mathbf{L}}=\mathbf{U}\mathbf{\Lambda}\mathbf{U}^\top$ where $\mathbf{\Lambda}=\text{Diag}(\lambda_1, \cdots, \lambda_{|\mathcal{V}|}), 0\le\lambda_1\leq \lambda_2\leq\cdots\leq \lambda_{|\mathcal{V}|}\le 2$, and $\mathbf{U}=[\mathbf{u}_1,\cdots, \mathbf{u}_{|\mathcal{V}|}]$ with $\mathbf{u}_i\in\mathbb{R}^{|\mathcal{V}|}$ being the eigenvector for eigenvalue $\lambda_i$. Then, given a graph signal $\mathbf{x}\in\mathbb{R}^{|\mathcal{V}|}$, its component on the $k$-th frequency is given by $\tilde{\mathbf{x}}_k=\mathbf{u}_k\mathbf{u}_k^\top\mathbf{x}$.


In the following section, we perform the frequency analysis on the user-user KNN graph and the item-item KNN graph separately. We construct them from the predictions of the teacher model. Specifically, before training, we calculate the Euclidean distances between each pair of users (or items) using the frozen teacher model and obtain the KNN graph. Thus, when performing our method on the user KNN graph (or the item KNN graph), the number of nodes is equal to the number of users (or items, respectively).

\subsection{Feature-based Knowledge Distillation}
Let $d^S$ and $d^T$ denote the dimensions of student and teacher features, respectively. 
In this work, we use $\mathbf{S}\in \mathbb{R}^{|\mathcal{V}|\times d^S}$ and $\mathbf{T}\in\mathbb{R}^{|\mathcal{V}|\times d^T}$ to represent all the features of the student model and the teacher model, respectively. Note that, when performing knowledge for users, we have $|\mathcal{V}|=|\mathcal{U}|$, otherwise $|\mathcal{V}|=|\mathcal{I}|$.

In feature-based knowledge distillation methods~\cite{kang2020rrd,kang2022personalized}, a projector is used to align the dimensionalities of the student model and the teacher model. Then, the student model is trained to minimize the MSE loss to make the student model mimic the teacher's features. Formally, the feature-based knowledge distillation loss is given by:
\begin{align*}
    \mathcal{L}_{FD}=\|Proj(\mathbf{S})-\mathbf{T}\|_F^2\,,
\end{align*}
where $Proj(\cdot)$ denotes the projector that aligns the dimensionalities of $\mathbf{S}$ and $\mathbf{T}$, and is usually a linear transformation.

\section{Methodology}

In Section~\ref{sec:anal}, we define the knowledge in features and show the difference between various types of knowledge. In Section~\ref{sec:reweight}, we propose a simple yet effective method to reweight the distillation loss of different knowledge. In Section~\ref{sec:light}, we further propose \ourmethod, a lightweight knowledge reweight method. The overall training process of \ourmethod is exhibited in Algorithm~\ref{alg} and the time complexity analysis is provided in Section~\ref{sec:complexity}. We further discuss the relationship of our method with FitNet in Section~\ref{sec:discuss}. Finally, we also discuss the prospect of extending our analytical approach to relation-based knowledge distillation methods in Appendix~\ref{sec:prospect}.

\subsection{Analysis of Knowledge in Feature-based Knowledge Distillation}\label{sec:anal}

First, we must define the knowledge in features to analyze the knowledge transfer process in feature-based knowledge distillation. Inspired by recent work in graph signal processing and graph collaborative filtering, we introduce the following definition.
\begin{definition}[Knowledge in features]
    Given the features $\mathbf{X}\in\{\mathbf{S}, \mathbf{T}\}$, the $k$-th type of knowledge contained in the features is defined as the  $k$-th frequency component of $\mathbf{X}$; that is
    \begin{align*}
        \tilde{\mathbf{X}}_k=\mathbf{u}_k\mathbf{u}_k^\top\mathbf{X}\,.
    \end{align*}
\end{definition}
Then, we have a total of $|\mathcal{V}|$ types of knowledge. As we have analyzed in previous sections, different types of knowledge indicate different degrees of commonality among nodes.

Next, we theoretically show that when minimizing feature-based knowledge distillation losses, the student model must learn all types of knowledge equally.

\begin{theorem}
    \label{theorem-freq-decomp}
    Consider the feature-based distillation loss:
    \begin{align*}
        \mathcal{L}_{FD} = \|Proj(\mathbf{S})-\mathbf{T}\|_F^2\,.
    \end{align*}
    Then $\mathcal{L}_{FD}$ can be decomposed into losses on all types of knowledge.
    Formally, 
    \begin{align*}
        \mathcal{L}_{FD} = \sum_{k=1}^{|\mathcal{V}|}\|Proj(\tilde{\mathbf{S}}_k)-\tilde{\mathbf{T}}_k\|_F^2\,.
    \end{align*}
\end{theorem}

The proof is presented in Appendix~\ref{app:theorem1}.
Theorem~\ref{theorem-freq-decomp} indicates that the feature-based distillation loss aims to force the student model to learn all types of knowledge with equal weights. However, it is difficult for the student to master all types of knowledge due to the vast difference in capacity between the student and the teacher. To verify this, we next report the distillation loss on various types of knowledge. Since there are too many types of knowledge to show separately, we evenly divide all knowledge into four consecutive groups, which are defined as follows:
\begin{align*}
    \begin{cases}
        \mathcal{S}_1=\{1,2,\cdots,\lfloor |\mathcal{V}|/4\rfloor\} & \text{Knowledge Group 1}\\
        \mathcal{S}_2=\{\lfloor |\mathcal{V}|/4\rfloor+1,\cdots,\lfloor |\mathcal{V}|/2\rfloor\} & \text{Knowledge Group 2}\\
        \mathcal{S}_3=\{\lfloor |\mathcal{V}|/2\rfloor+1,\cdots,\lfloor 3|\mathcal{V}|/4\rfloor\} & \text{Knowledge Group 3}\\
        \mathcal{S}_4=\{\lfloor 3|\mathcal{V}|/4\rfloor+1,\cdots,|\mathcal{V}|\} & \text{Knowledge Group 4}\\
    \end{cases}
\end{align*}
Then, the distillation loss on the $i$-th knowledge group is given by
\begin{align*}
    \mathcal{L}_{\mathcal{S}_i}=\sum_{k\in\mathcal{S}_i}\|Proj(\tilde{\mathbf{S}}_k)-\tilde{\mathbf{T}}_k\|_F^2, \quad i=1,2,3,4.
\end{align*}
As a special case of Theorem~\ref{theorem-freq-decomp}, we have
\begin{align}\label{eq:loss_4group}
    \mathcal{L}_{FD}=\sum_{i=1}^4\mathcal{L}_{S_i}.
\end{align}

\begin{figure}[!t]
\centering
  \includegraphics[width=\linewidth]{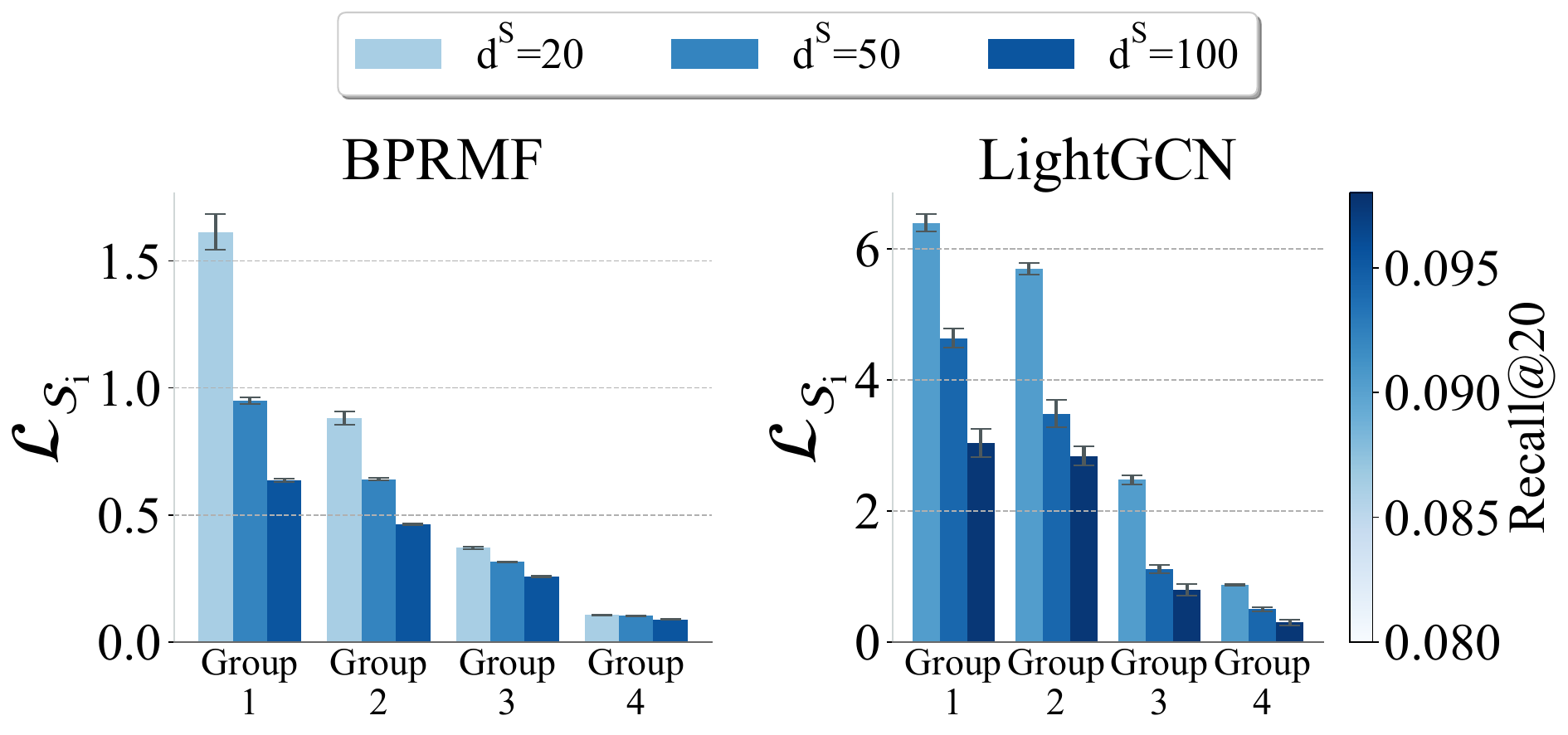}
  \caption{Feature-based distillation loss on different knowledge groups. Colors indicate Recall@20 of these students. The two backbones used in the experiments are BPRMF (left) and LightGCN (right), respectively.}
  \label{fig:fd_loss_diffs}
 \vspace{-4pt}
\end{figure}

Figure~\ref{fig:fd_loss_diffs} shows the feature-based distillation loss on different knowledge groups. The experiments are conducted using students of different dimensionalities, and we also report their Recall@20. From the results, we find that: (1) All students have a greater distillation loss on knowledge groups representing low frequencies than on high frequencies. (2) The smaller the capacity of the student model, the greater the losses on knowledge groups representing low frequencies.

\subsection{Knowledge Reweighting}\label{sec:reweight}

Our findings in the previous section suggest that during the knowledge distillation process, knowledge representing low frequencies fails to be transferred to student models of low capacity. Unfortunately, our experiment next illustrates that this neglected knowledge is very important. Specifically, we first rewrite the feature-based distillation loss in Eq.(\ref{eq:loss_4group}) as:
\begin{align}\label{eq:rewrite}
    \mathcal{L}_{FD}=\sum_{i=1}^4w_i\mathcal{L}_{\mathcal{S}_i}.
\end{align}

Then, we investigate the importance of the $i$-th knowledge group by setting $w_i=0$ and $w_j=1,\forall j\neq i$. The results are shown in Table~\ref{tab:wo_diffk}. We observe that removing knowledge representing low frequencies leads to a significant decrease in recommendation performance, which validates our claim that the neglected knowledge is critical.

\begin{table}[!t]
    \caption{Performance of the student after removing various types of knowledge ($w_1$ corresponds to low frequency).}
    \label{tab:wo_diffk}
    \centering
    \begin{tabular}{cccc} \toprule
    Backbone & Method & Recall@20 & NDCG@20\\
    \hline
    \multirow{5}{*}{BPRMF} & Original & 0.0862 & 0.0432\\
     & $w_1=0$ & 0.0657 & 0.0317\\
     & $w_2=0$ & 0.0799 & 0.0398\\
     & $w_3=0$ & 0.0842 & 0.0417\\
     & $w_4=0$ & 0.0855 & 0.0427\\
     \hline
    \multirow{5}{*}{LightGCN} & Original & 0.0903 & 0.0459\\
     & $w_1=0$ & 0.0785 & 0.0352\\
     & $w_2=0$ & 0.0800 & 0.0381\\
     & $w_3=0$ & 0.0873 & 0.0429\\
     & $w_4=0$ & 0.0892 & 0.0450\\
    \bottomrule
    \end{tabular}
\end{table}

\begin{figure}[!t]
    \vspace{-2pt}
  \centering
  \includegraphics[width=\linewidth]{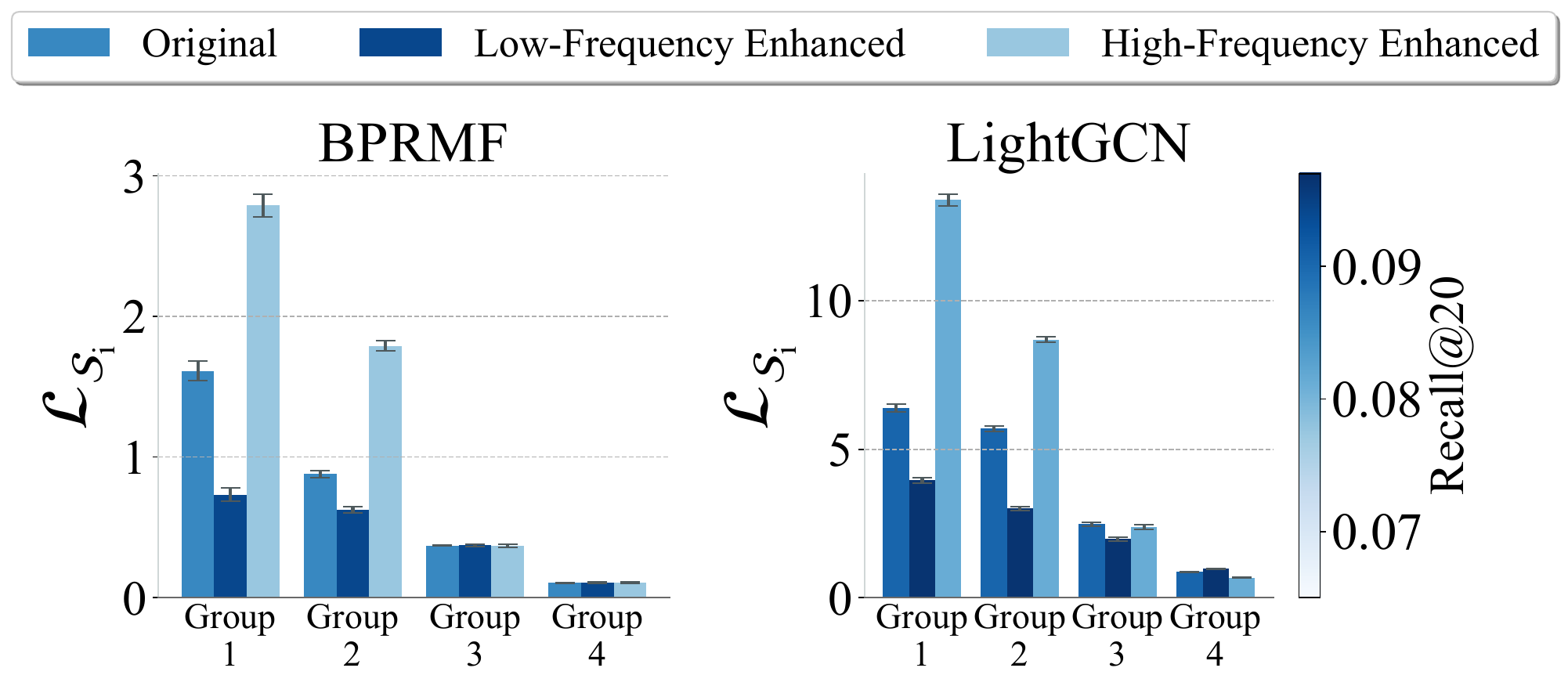}
  \caption{Feature-based distillation loss on different knowledge groups for different weight allocation schemes.}
  \label{fig:fd_loss_diff_weight}
    \vspace{-7pt}
\end{figure}

\begin{table*}[!t]
    \caption{Dimensionalities of teachers and students.}
    \label{tab:dim}
    \vspace{-0.5em}
    \centering
    \begin{tabular}{c|ccc|ccc|ccc} \toprule
        \multirow{2}{*}{Method} & \multicolumn{3}{c|}{CiteULike} & \multicolumn{3}{c|}{Gowalla} & \multicolumn{3}{c}{Yelp}\\
         & BPRMF & LightGCN & SimpleX & BPRMF & LightGCN & SimpleX & BPRMF & LightGCN & SimpleX\\
        \midrule
        Teacher & 400 & 2000 & 400 & 300 & 2000 & 1000 & 300 & 1000 & 500\\
        Student & 20 & 20 & 20 & 20 & 20 & 20 & 20 & 20 & 20\\
        \bottomrule
    \end{tabular}
\end{table*}

The findings above emphasize that students should pay more attention to knowledge representing low frequencies. We propose that this can be achieved by reweighting the losses on different knowledge groups. Specifically, we should set higher weights for knowledge groups representing low frequencies than others. To verify the effectiveness of this weight allocation scheme, we propose three weight allocation scheme which set different values for $w_i$ in Eq.(\ref{eq:rewrite}): (1) \textit{Original} sets $w_1=w_2=w_3=w_4=1$; (2) \textit{Low-Frequency Enhanced} sets $w_1=1.0, w_2=0.75, w_3=0.5, w_4=0.25$; (3) \textit{High-Frequency Enhanced} sets $w_1=0.25, w_2=0.5, w_3=0.75, w_4=1.0$.

As shown in Figure~\ref{fig:fd_loss_diff_weight}, we significantly reduce the student's loss on the low-frequency knowledge groups by setting higher weights for knowledge groups representing low frequencies. Since the student model can easily mimic the teacher's features on high-frequency knowledge groups, the distillation losses on the high-frequency knowledge remain almost unchanged. Moreover, by comparing the recommendation performance of these three weight allocation schemes, we observe that reducing the losses on low-frequency knowledge groups helps improve the student model's performance.

With the above analysis, we propose to assign larger weights to the knowledge representing low frequencies than the knowledge representing high frequencies. Formally, 
\begin{align}
\label{eq:loss}
&\mathcal{L}_{KRD}=\sum_{k=1}^{|\mathcal{V}|} g(\lambda_k)\|Proj(\tilde{\mathbf{S}}_k)-\tilde{\mathbf{T}}_k\|_F^2,\\
&\text{s.t. } g(x_1) \geq g(x_2), \forall x_1\le x_2.
\end{align}
where $\lambda_k$ denotes the frequency of the $k$-th knowledge (i.e., $k$-th eigenvalue of the graph Laplacian matrix). Decreasing function $g(\cdot)$ is defined on the frequencies and determines the weight of the loss on each knowledge. Then, all we need to do is to design the decreasing function $g(\cdot)$, which will be discussed in the next section.

\subsection{Lightweight Knowledge Reweighting Method}\label{sec:light}

Although the above method allows us to specify the weight for each knowledge, direct computation requires computing each knowledge $\tilde{\mathbf{X}}_k$, which also requires a complete eigendecomposition of the graph Laplacian matrix to obtain $\mathbf{u}_k$. However, this is usually impractical because there are many users and items in real recommendation scenarios. In this section, we propose \ourmethod, a lightweight weight knowledge reweighting method that defines graph filters and performs graph filtering on the features to assign weights to the losses on different knowledge indirectly.

First, we demonstrate that computing feature-based distillation loss using filtered features is equivalent to reweighting losses on different knowledge.

\begin{theorem}
    \label{theorem-filter}
    Given a graph filter $\mathcal{H}(\tilde{\mathbf{L}})$ defined as follows:
    \begin{align}\label{eq:HL}
        \mathcal{H}(\tilde{\mathbf{L}})=\mathbf{U}\text{Diag}(h(\lambda_1), h(\lambda_2), \cdots, h(\lambda_{|\mathcal{V}|})\mathbf{U}^\top
    \end{align}
    where $\tilde{\mathbf{L}}$ is the graph Laplacian matrix. Then, computing the feature-based distillation loss using the features filtered by $\mathcal{H}(\tilde{\mathbf{L}})$ is equivalent to setting weights $g(\lambda_i)=h^2(\lambda_i)$ in Eq.(\ref{eq:loss}).
    
    Formally, we have
    \begin{align*}
        \|\mathcal{H}(\tilde{\mathbf{L}})Proj(\mathbf{S})-\mathcal{H}(\tilde{\mathbf{L}})\mathbf{T}\|_F^2=\sum_{k=1}^{|\mathcal{V}|} h^2(\lambda_k)\|(Proj(\tilde{\mathbf{S}}_k)-\tilde{\mathbf{T}}_k)\|_F^2.
    \end{align*}
\end{theorem}

The proof is provided in Appendix~\ref{app:theorem2}. Based on Theorem~\ref{theorem-filter}, we have now transformed the task of designing the decreasing function $g(\cdot)$ presented in the previous section into designing the decreasing function $h(\cdot)$. However, the direct computation of $\mathcal{H}(\tilde{\mathbf{L}})$ using Eq.(\ref{eq:HL}) still requires an eigendecomposition of the graph Laplacian matrix to obtain $\mathbf{U}$. Inspired by work in graph signal processing, we propose to solve this problem by choosing $h(\lambda)$ as a polynomial function of $\lambda$, which has the following general formulation:
\begin{align}
h(\lambda)=\sum_{k=0}^K\theta_k\lambda^k\,,\label{eq:h_lambda}
\end{align}
where $K$ defines the order of the polynomial function and $\theta_k$ represents the coefficient of the $k$-th term. Following LightGCN~\cite{he2020lightgcn}, we regard $\theta_k$ as hyperparameters.
The specific choices of $h(\lambda)$ are shown in the implementation part of Section~\ref{subsec:exp-setting}, 

Then, the computation of $\mathcal{H}(\tilde{\mathbf{L}})$ can be simply implemented by:
\begin{align}\label{eq:our_filter}
    \mathcal{H}(\tilde{\mathbf{L}})=\sum_{k=0}^K\theta_k\tilde{\mathbf{L}}^k.
\end{align}

Consequently, the lightweight knowledge-reweighted feature-based distillation loss is given by
\begin{align}\label{eq:final}
    \mathcal{L}_{FreqD}=\|\mathcal{H}(\tilde{\mathbf{L}})Proj(\mathbf{S})-\mathcal{H}(\tilde{\mathbf{L}})\mathbf{T}\|_F^2.
\end{align}

Finally, the total loss for training the student is given by
\begin{align}\label{eq:final_loss}
    \mathcal{L}=\mathcal{L}_{Base}+\beta\cdot \mathcal{L}_{FreqD}\,,
\end{align}
where $\mathcal{L}_{Base}$ is the loss function of the base recommendation model, such as BPR loss. $\beta$ is the hyperparameter that balances the two losses.

The pseudocode of the entire training process is presented in Algorithm~\ref{alg}.

\begin{algorithm*}
    \caption{\ourmethod}\label{alg}
    \SetKwInOut{Input}{Input}
    \SetKwInOut{Output}{Output}
    \Input{Training data $\mathcal{R}$; Pretrained teacher model $T$.}
    \Output{Student model $S$.}
    Randomly initialize the Student model $S$ and the projector $Proj(\cdot)$.\\
    Construct the user-item bipartite graph and obtain the asymmetric normalized graph Laplacian matrix using $\mathcal{R}$.\\
    Compute the graph filter $\mathcal{H}(\tilde{\mathbf{L}})$ using Eq.(\ref{eq:our_filter}).\\
    \While{not convergence}{
        \For{each batch $\mathcal{B}\in \mathcal{R}$}{
            Compute $\mathcal{L}_{Base}$.\\
            Compute filtered student features $\mathcal{H}(\tilde{\mathbf{L}})\mathbf{S}$ and teacher features $\mathcal{H}(\tilde{\mathbf{L}})\mathbf{T}$.\\
            Obtain post-projector features $\mathcal{H}(\tilde{\mathbf{L}})Proj(\mathbf{S})$.\\
            Compute $\mathcal{L}_{FreqD}$ according to Eq.(\ref{eq:final}).\\
            Compute $\mathcal{L}=\mathcal{L}_{Base}+\beta\cdot \mathcal{L}_{FreqD}$.\\
            Update the Student model and the projector by minimizing $\mathcal{L}$.
        }
    }
\end{algorithm*}

\subsection{Time Complexity}\label{sec:complexity}
The proposed \ourmethod is highly efficient. The only additional operation that our method performs over FitNet is graph filtering. Thanks to the support for sparse matrix multiplication in various machine learning libraries, the complexity of this filtering operation is only $\mathcal{O}(|\mathcal{R}|(d^S+d^T))$. Given batch size $|\mathcal{B}|$, since the complexity of FitNet is $\mathcal{O}(|\mathcal{B}|d^Sd^T)$, the complexity of our method is $\mathcal{O}(|\mathcal{B}|d^Sd^T+|\mathcal{R}|(d^S+d^T))$. In contrast, the complexity of DE using $K$ projectors is $\mathcal{O}(K|\mathcal{B}|d^Sd^T)$, and the required training time is usually larger than our method.

\subsection{Further Discussion}\label{sec:discuss}

\textbf{Relation to FitNet.}
FitNet~\cite{romero2014fitnets} is a widely used method for feature-based KD. It simply uses conventional feature-based knowledge distillation loss. It can regarded be as a special case of our method. Specifically, we can rewrite Eq.(\ref{eq:final}) as follows:
\begin{align}
\mathcal{L}_{FreqD}&=\sum_{k=1}^{|\mathcal{V}|} h^2(\lambda_k)\|(Proj(\tilde{\mathbf{S}}_k)-\tilde{\mathbf{T}}_k)\|_F^2\\
&=\left\|\left(\sum_{k=0}^K\theta_k\tilde{\mathbf{L}}^k\right)Proj(\textbf S)-\left(\sum_{k=0}^K\theta_k\tilde{\mathbf{L}}^k\right)\textbf T\right\|_F^2.
\end{align}

It follows that FitNet can be regarded as a 0-order special case of our approach. Specifically, by setting the hyperparameter $K$ to $0$, we can exactly recover FitNet. It also means that all frequency components are weighted equally in FitNet, i.e., $h^2(\lambda_k), k=1,\cdots,|\mathcal V|$ is constant. Combined with our previous analysis, it can be assumed that we have improved FitNet based on the fact that ``low frequency'' is more important in the field of recommender systems.

\begin{table}
    \caption{Statistics of the preprocessed datasets.}
    \vspace{-0.5em}
    \centering
    \begin{tabular}{ccccc}\toprule
         Dataset & \#Users & \#Items & \#Interactions & \#Sparsity\\
         \midrule
         \texttt{CiteULike} & 5,219 & 25,181 & 125,580 & 99.89\%\\
         \texttt{Gowalla} & 29,858 & 40,981 & 1,027,370 & 99.92\%\\
         \texttt{Yelp2018} & 41,801 & 26,512 & 1,022,604 & 99.91\%\\
         \bottomrule
    \end{tabular}
    \label{tab:dataset}
\end{table}

\begin{table*}[!t]
    \caption{Recommendation performance. The best results are in boldface, and the best baselines are underlined. \textit{Improv.b} denotes the relative improvement of \ourmethod over the best baseline. A paired t-test is performed over 5 independent runs for evaluating $p$-value ($\leq 0.05$ indicates statistical significance).}
    \label{tab:all}
    \vspace{-0.5em}
    \centering
    \resizebox{\linewidth}{!}{
    \begin{tabular}{cc|cccc|cccc|cccc} \toprule
        \multirow{2}{*}{Backbone} & \multirow{2}{*}{Method} & \multicolumn{4}{c|}{CiteULike} & \multicolumn{4}{c|}{Gowalla} & \multicolumn{4}{c}{Yelp}\\
         & & R@10 & N@10 & R@20 & N@20 & R@10 & N@10 & R@20 & N@20 & R@10 & N@10 & R@20 & N@20\\
        \midrule
        \multirow{10}{*}{BPRMF} & Teacher & 0.0283 & 0.0155 & 0.0442 & 0.0198 & 0.1088 & 0.0907 & 0.1544 & 0.1053 & 0.0394 & 0.0253 & 0.0660 & 0.0339\\
         & Student & 0.0177 & 0.0098 & 0.0284 & 0.0128 & 0.0946 & 0.0820 & 0.1329 & 0.0939 & 0.0348 & 0.0222 & 0.0586 & 0.0299\\
         \cline{2-14}
         & CD & 0.0239 & 0.0131 & 0.0347 & 0.0158 & 0.0979 & 0.0855 & 0.1389 & 0.0977 & \underline{0.0370} & \underline{0.0236} & 0.0608 & 0.0310\\
         & RRD & \underline{0.0251} & \underline{0.0135} & \underline{0.0362} & \underline{0.0169} & 0.0977 & \underline{0.0861} & 0.1395 & \underline{0.0987} & 0.0362 & 0.0230 & \underline{0.0626} & \underline{0.0319}\\
         & UnKD & 0.0242 & 0.0129 & 0.0348 & 0.0160 & \underline{0.0981} & 0.0860 & \underline{0.1401} & 0.0979 & 0.0366 & 0.0228 & 0.0601 & 0.0312\\
         & HTD & 0.0235 & 0.0127 & 0.0337 & 0.0152 & 0.0975 & 0.0857 & 0.1379 & 0.0970 & 0.0360 & 0.0231 & 0.0600 & 0.0309\\
         & DE & 0.0212 & 0.0119 & 0.0331 & 0.0147 & 0.0969 & 0.0847 & 0.1372 & 0.0971 & 0.0355 & 0.0227 & 0.0598 & 0.0305\\
         & FitNet & 0.0199 & 0.0108 & 0.0326 & 0.0140 & 0.0963 & 0.0841 & 0.1352 & 0.0967 & 0.0355 & 0.0226 & 0.0599 & 0.0301\\
         & \ourmethod & \textbf{0.0282} & \textbf{0.0154} & \textbf{0.0428} & \textbf{0.0194} & \textbf{0.1056} & \textbf{0.0902} & \textbf{0.1494} & \textbf{0.1039} & \textbf{0.0398} & \textbf{0.0257} & \textbf{0.0669} & \textbf{0.0344}\\
         \cline{2-14}
         & \textit{Improv.b} & 12.35\% & 14.07\% & 18.23\% & 14.79\% & 7.64\% & 4.76\% & 6.64\% & 5.26\% & 7.57\% & 8.90\% & 6.87\% & 7.84\%\\
         & \textit{p-value} & 7.24e-5 & 4.18e-4 & 3.92e-5 & 1.71e-4 & 3.88e-4 & 9.91e-3 & 4.07e-3 & 8.62e-3 & 3.72e-4 & 8.28e-5 & 3.88e-4 & 7.62e-4\\
        \midrule
        \midrule
         \multirow{10}{*}{LightGCN} & Teacher & 0.0296 & 0.0160 & 0.0461 & 0.0205 & 0.1236 & 0.1035 & 0.1730 & 0.1190 & 0.0432 & 0.0276 & 0.0716 & 0.0367\\
         & Student & 0.0215 & 0.0113 & 0.0344 & 0.0148 & 0.1098 & 0.0928 & 0.1550 & 0.1069 & 0.0363 & 0.0235 & 0.0621 & 0.0308\\
         \cline{2-14}
         & CD & 0.0234 & \underline{0.0125} & 0.0354 & \underline{0.0161} & 0.1132 & 0.0951 & 0.1592 & 0.1100 & 0.0385 & \underline{0.0247} & 0.0669 & \underline{0.0342}\\
         & RRD & \underline{0.0247} & 0.0125 & \underline{0.0359} & 0.0158 & \underline{0.1142} & \underline{0.0969} & \underline{0.1627} & 0.1109 & \underline{0.0391} & 0.0245 & \underline{0.0671} & 0.0338\\
         & UnKD & 0.0233 & 0.0122 & 0.0353 & 0.0157 & 0.1138 & 0.0967 & 0.1622 & \underline{0.1112} & 0.0390 & 0.0242 & 0.0658 & 0.0329\\
         & HTD & 0.0231 & 0.0120 & 0.0352 & 0.0154 & 0.1139 & 0.0951 & 0.1589 & 0.1102 & 0.0375 & 0.0245 & 0.0653 & 0.0330\\
         & DE & 0.0229 & 0.0119 & 0.0349 & 0.0155 & 0.1111 & 0.0949 & 0.1578 & 0.1097 & 0.0369 & 0.0242 & 0.0641 & 0.0319\\
         & FitNet & 0.0220 & 0.0118 & 0.0350 & 0.0151 & 0.1103 & 0.0939 & 0.1571 & 0.1088 & 0.0368 & 0.0238 & 0.0633 & 0.0321\\
         & \ourmethod & \textbf{0.0263} & \textbf{0.0134} & \textbf{0.0384} & \textbf{0.0171} & \textbf{0.1198} & \textbf{0.1020} & \textbf{0.1712} & \textbf{0.1175} & \textbf{0.0411} & \textbf{0.0257} & \textbf{0.0702} & \textbf{0.0359}\\
         \cline{2-14}
         & \textit{Improv.b} & 6.48\% & 7.20\% & 6.96\% & 6.21\% & 4.90\% & 5.26 & 5.22\% & 5.67\% & 5.12\% & 4.05\% & 4.62\% & 4.97\%\\
         & \textit{p-value} & 2.24e-3 & 8.72e-4 & 1.26e-3 & 3.88e-3 & 1.27e-3 & 5.82e-4 & 2.97e-3 & 3.18e-3 & 1.02e-5 & 7.17e-3 & 3.82e-4& 6.97e-4\\
         \midrule
         \midrule
         \multirow{10}{*}{SimpleX} & Teacher & 0.0343 & 0.0191 & 0.0508 & 0.0236 & 0.1184 & 0.0943 & 0.1750 & 0.1122 & 0.0469 & 0.0303 & 0.0778 & 0.0402\\
         & Student & 0.0290 & 0.0162 & 0.0426 & 0.0199 & 0.1046 & 0.0855 & 0.1524 & 0.1006 & 0.0382 & 0.0241 & 0.0658 & 0.0330\\
         \cline{2-14}
         & CD & 0.0319 & 0.0170 & 0.0462 & 0.0213 & 0.1092 & \underline{0.0902} & \underline{0.1621} & \underline{0.1071} & 0.0425 & 0.0271 & 0.0702 & 0.0361\\
         & RRD & \underline{0.0328} & \underline{0.0175} & \underline{0.0478} & 0.0219 & \underline{0.1101} & 0.0893 & 0.1615 & 0.1059 & \underline{0.0435} & 0.0279 & \underline{0.0717} & \underline{0.0371}\\
         & UnKD & 0.0326 & 0.0175 & 0.0459 & \underline{0.0223} & 0.1081 & 0.0887 & 0.1607 & 0.1061 & 0.0431 & \underline{0.0287} & 0.0713 & 0.0369\\
         & HTD & 0.0321 & 0.0172 & 0.0462 & 0.0215 & 0.1089 & 0.0883 & 0.1599 & 0.1057 & 0.0427 & 0.0278 & 0.0690 & 0.0358\\
         & DE & 0.0313 & 0.0165 & 0.0449 & 0.0203 & 0.1073 & 0.0869 & 0.1567 & 0.1042 & 0.0409 & 0.0269 & 0.0671 & 0.0349\\
         & FitNet & 0.0305 & 0.0167 & 0.0434 & 0.0205 & 0.1064 & 0.0862 & 0.1558 & 0.1031 & 0.0398 & 0.0268 & 0.0667 & 0.0341\\
         & \ourmethod & \textbf{0.0341} & \textbf{0.0187} & \textbf{0.0507} & \textbf{0.0236} & \textbf{0.1171} & \textbf{0.0939} & \textbf{0.1739} & \textbf{0.1110} & \textbf{0.0461} & \textbf{0.0299} & \textbf{0.0758} & \textbf{0.0392}\\
         \cline{2-14}
         & \textit{Improv.b} & 3.96\% & 6.86\% & 6.07\% & 5.83\% & 6.36\% & 4.10\% & 7.28\% & 3.64\% & 5.98\% & 4.18\% & 5.72\% & 5.66\%\\
         & \textit{p-value} & 1.2e-2 & 7.92e-5 & 2.77e-4 & 3.28e-3 & 4.28e-5 & 2.77e-3 & 9.62e-5 & 1.83e-2 & 3.82e-3 & 7.28e-3 & 9.52e-4 & 2.44e-5\\
        \bottomrule
    \end{tabular}
    }
\end{table*}

\section{Experiments}
In this section, we conduct experiments on three public datasets using three backbones to validate the effectiveness of \ourmethod. The overall performance comparison and ablation study are presented in Section~\ref{sec:result}. The analysis of training efficiency and inference efficiency are presented in Section~\ref{sec:eff} and Section~\ref{sec:inference_eff}, respectively. In Section~\ref{sec:h_lambda}, we provide the performance when $h(\lambda)$ is set to a quadratic function form. Some key hyperparameter analysis is provided in Appendix~\ref{sec:hyper}.

\subsection{Experimental Settings}\label{subsec:exp-setting}

\noindent\textbf{Datasets.}
We conduct experiments on three public datasets, including \textbf{CiteULike}\footnote{\url{https://github.com/changun/CollMetric/tree/master/citeulike-t}}~\cite{wang2013collaborative,kang2022personalized,kang2021topology}, \textbf{Gowalla}\footnote{\url{http://dawenl.github.io/data/gowalla pro.zip}}~\cite{cho2011friendship,tang2018ranking,lee2019collaborative}, and \textbf{Yelp2018}\footnote{\url{https://github.com/hexiangnan/sigir16-eals}}~\cite{lee2019collaborative,kweon2021bidirectional}. We split training and test dataset following the previous method~\cite{xu2023stablegcn}. Specifically, we filter out users and items with less than 10 interactions and then split the rest chronologically into training, validation, and test sets in an 8:1:1 ratio. The statistics of the preprocessed datasets are summarized in Table~\ref{tab:dataset}.

\noindent\textbf{Evaluation Protocols.}
Per the custom, we adopt the full-ranking evaluation to achieve an unbiased evaluation. To evaluate the performance of top-$N$ recommendation, we employ Recall (Recall@$N$) and normalized discounted cumulative gain (NDCG@$N$) and report the results for $N \in \{10, 20\}$. We conduct five independent runs for each configuration and report the averaged results.

\noindent\textbf{Baselines.}
We compare our method with all three types of knowledge distillation methods:
\begin{itemize}[leftmargin=*]
\item For \textit{response-based} methods, we choose \textbf{CD}~\cite{lee2019collaborative}, \textbf{RRD}~\cite{kang2020rrd}, and \textbf{UnKD}\cite{chen2023unbiased}. They all use the teacher's logits to provide additional supervision to the student.
\item For \textit{relation-based} method, we select \textbf{HTD}~\cite{kang2021topology}. It uses the selector network to partition the samples and distill the sample relations hierarchically.
\item For \textit{feature-based} methods, we select \textbf{DE}~\cite{kang2020rrd} and \textbf{FitNet}~\cite{romero2014fitnets}. They belong to the same category as our approach, and all use the teacher model's features.
\end{itemize}

\noindent\textbf{Backbones.}
We refer to previous works~\cite{chen2023unbiased,kang2020rrd,kang2021topology}, and use BPRMF~\cite{rendle2012bpr} and LightGCN~\cite{he2020lightgcn}.  We also add SimpleX~\cite{mao2021simplex} as a new backbone.

\noindent\textbf{Teacher/Student.}
Our work focuses on the scenario where teachers and students have the same backbone. For each backbone, we increase the model size until the recommendation performance is no longer improved and adopt the model with the best performance as the teacher model. We set the representation dimensionality of the student model to 20. The dimensionalities of the teachers and the students are summarized in Table~\ref{tab:dim}.

\noindent\textbf{Implementation Details.}
We implement all the methods with PyTorch and use Adam as the optimizer in all experiments. For our method, the weight decay is selected from \{1e-3, 1e-4, 1e-5, 0\}. The search space of the learning rate is \{1e-3, 1e-4\}. $\beta$ is selected from \{0.01, 0.05, 0.1, 0.5\}. We set the total number of training epochs as 1000. We also conduct early stopping according to the NDCG@20 on the validation set and stop training when the NDCG@20 doesn't increase for 30 consecutive epochs. All hyperparameters of the compared baselines are tuned to ensure optimal performance. Before graph filtering on the KNN graph, we perform edge dropout, following the practice in LightGCN~\cite{he2020lightgcn} to alleviate overfitting.

In the subsequent sections, unless otherwise noted, we set $h(\lambda)$ in Eq.(\ref{eq:h_lambda}) to a decreasing linear function. Formally, we set 
\begin{align}
    h(\lambda)=1-\alpha\lambda\label{eq:linear},
\end{align}
where $0\leq \alpha\leq 0.5$ is a hyperparameter. We have found it to be well-behaved while being very simple and efficient. Moreover, in Section~\ref{sec:h_lambda}, we provide the performance when $h(\lambda)$ is set to a quadratic function form for further verification.

\begin{table*}[!ht]
    \caption{The comparison of the training time (seconds) per epoch.}
    \label{tab:train_time}
    \vspace{-0.5em}
    \centering
    \begin{tabular}{c|ccc|ccc|ccc} \toprule
        \multirow{2}{*}{Method} & \multicolumn{3}{c|}{CiteULike} & \multicolumn{3}{c|}{Gowalla} & \multicolumn{3}{c}{Yelp}\\
         & BPRMF & LightGCN & SimpleX & BPRMF & LightGCN & SimpleX & BPRMF & LightGCN & SimpleX\\
        \midrule
        Student & 3.58 & 5.15 & 8.89 & 28.60 & 50.11 & 62.42 & 28.94 & 44.43 & 55.90\\
        \midrule
        CD & 23.14 & 28.87 & 37.99 & 162.23 & 199.97 & 273.20 & 228.81 & 258.38 & 377.72\\
        RRD & 22.57 & 24.61 & 33.35 & 158.70 & 193.40 & 265.18 & 204.58 & 246.63 & 351.77\\
        UnKD & 25.15 & 34.52 & 42.48 & 188.32 & 215.33 & 288.72 & 300.15 & 288.84 & 383.12\\
        HTD & 10.11 & 25.62 & 38.77 & 65.01 & 210.37 & 277.97 & 78.67 & 292.79 & 278.80\\
        DE & 6.17 & 15.26 & 32.13 & 62.43 & 188.53 & 269.73 & 52.36 & 199.80 & 244.79\\
        FitNet & 4.63 & 8.46 & 15.09 & 39.68 & 76.65 & 120.67 & 38.69 & 63.56 & 113.62\\
        \ourmethod & 5.92 & 10.97 & 23.52 & 49.79 & 116.48 & 193.18 & 47.46 & 105.88 & 188.69\\
        \bottomrule
    \end{tabular}
\end{table*}

\begin{table*}[!ht]
    \caption{The comparison of GPU Memory (GB) required by our method and comparison methods.}
    \label{tab:train_gpu}
    \vspace{-0.5em}
    \centering
    \begin{tabular}{c|ccc|ccc|ccc} \toprule
        \multirow{2}{*}{Method} & \multicolumn{3}{c|}{CiteULike} & \multicolumn{3}{c|}{Gowalla} & \multicolumn{3}{c}{Yelp}\\
         & BPRMF & LightGCN & SimpleX & BPRMF & LightGCN & SimpleX & BPRMF & LightGCN & SimpleX\\
        \midrule
        Student & 0.39 & 0.60 & 0.72 & 0.45 & 1.08 & 1.10 & 0.45 & 0.81 & 1.36\\
        \midrule
        CD & 1.07 & 2.52 & 1.21 & 5.49 & 8.81 & 19.01 & 4.99 & 7.02 & 13.10\\
        RRD & 0.92 & 2.42 & 1.20 & 5.23 & 8.64 & 18.89 & 4.85 & 6.30 & 12.62\\
        UnKD & 1.22 & 2.71 & 1.52 & 5.50 & 8.90 & 19.77 & 5.13 & 7.97 & 13.89\\
        HTD & 2.47 & 10.46 & 8.47 & 5.88 & 12.46 &  24.97 & 3.53 & 9.00 & 14.87\\
        DE & 0.85 & 3.54 & 3.28 & 1.93 & 5.68 & 8.81 & 0.95 & 2.69 & 5.41\\
        FitNet & 0.84 & 2.39 & 1.60 & 1.16 & 4.62 & 5.12 & 0.92 & 2.70 & 3.06\\
        \ourmethod & 0.84 & 2.39 & 1.60 & 1.18 & 4.62 & 5.12 & 0.95 & 2.70 & 3.06\\
        \bottomrule
    \end{tabular}
\end{table*}

\subsection{Performance Comparison}\label{sec:result}

\textbf{Overall Performance.} The results shown in
Table~\ref{tab:all} indicate that:

 \noindent\textbf{(\romannumeral1)} \ourmethod outperforms all baselines significantly on all datasets and achieves remarkable improvements over the best baseline for different backbones, consolidating the effectiveness of \ourmethod.

 \noindent\textbf{(\romannumeral2)} In most cases, response-based knowledge distillation methods are better than other baseline methods, including feature-based methods and relation-based methods. This is at odds with the excellent performance of feature-based distillation methods in CV and NLP, reflecting the importance of a deeper understanding of feature-based distillation methods for recommender systems.
 
 \noindent\textbf{(\romannumeral3)} On all three datasets, \ourmethod shows comparable performance to the teacher, making it possible to achieve inference efficiency without sacrificing recommendation accuracy.

\textbf{Ablation Study.}
Since FitNet can be regarded as a special case of our method obtained by equally weighting losses on all types of knowledge, here we serve as an ablation experiment by comparing our method with FitNet. By comparing the performance of FitNet and \ourmethod in Table~\ref{tab:all}, we find that although \ourmethod is simple, it significantly outperforms FitNet on all datasets and all backbones. This suggests that by decoupling the knowledge contained in the features and controlling the weights of each type of knowledge separately, we can effectively facilitate the learning of more important knowledge by the student model and consequently gain improved results. It also shows that the conclusions we get from decoupling the various types of knowledge in features and analyzing their performance in feature-based distillation are useful for designing practical methods.
 We believe that our design concept will be useful for other knowledge distillation methods in the field of recommender systems.

\begin{table*}[!ht]
    \caption{Inference efficiency of \ourmethod with different student model sizes. $\phi$ denotes the ratio of the student's dimensionality to the teacher's. Time (seconds) indicates the inference latency. \#Params denotes the number of parameters. R-R@10 denotes the ratio of the student model's R@10 gained by \ourmethod to the teacher model's R@10.}
    \label{tab:effi}
    \vspace{-0.5em}
    \centering
    \begin{tabular}{cc|ccc|ccc|ccc} \toprule
        \multirow{2}{*}{Backbone} & \multirow{2}{*}{$\phi$} & \multicolumn{3}{c|}{CiteULike} & \multicolumn{3}{c|}{Gowalla} & \multicolumn{3}{c}{Yelp}\\
         & & Time (s) & \#Params. & R-R@10 & Time (s) & \#Params. & R-R@10 & Time (s) & \#Params. & R-R@10\\
        \midrule
        \multirow{3}{*}{BPRMF} & 1.0 & 23.79 & 11.60M & 1.21 & 125.52 & 20.27M & 1.16 & 172.13 & 19.54M & 1.17\\
         & 0.5 & 17.66 & 5.80M & 1.15 & 97.46 & 10.13M & 1.08 & 144.31& 9.76M & 1.10\\
         & 0.1 & 4.37 & 1.16M & 1.03 & 31.22 & 2.03M & 1.00 & 47.92 & 1.95M & 1.03\\
        \midrule
        \midrule
        \multirow{3}{*}{LightGCN} & 1.0 & 41.36 & 57.98M & 1.16 & 287.62 & 135.11M & 1.20 & 298.48 & 130.30M & 1.14\\
         & 0.1 & 27.20& 5.80M & 1.09 & 129.37 & 13.51M & 1.12 & 155.82 & 13.03M & 1.07\\
         & 0.01 & 9.13 & 0.58M & 0.89 & 61.43 & 1.35M & 0.96 & 79.73 & 1.30M & 0.91\\
        \midrule
        \midrule
        \multirow{3}{*}{SimpleX} & 1.0 & 30.29 & 14.73M & 1.17 & 169.44 & 68.51M & 1.19 & 192.76 & 32.81M & 1.16\\
         & 0.5 & 22.41 & 7.31M & 1.12 & 127.82 & 34.02M & 1.10& 148.97 & 16.35M & 1.10\\
         & 0.1 & 7.81 & 1.45M & 1.01 & 49.38 & 6.77M & 1.00 & 58.29 & 3.26M & 1.00\\
        \bottomrule
    \end{tabular}
\end{table*}

\subsection{Training Efficiency}\label{sec:eff}

In this section, we report the training efficiency of our method and comparison methods. All results are obtained by testing with PyTorch on GeForce RTX 3090 GPU.

In \ourmethod, we only need to perform graph filtering on features once before proceeding with regular feature-based knowledge distillation. Moreover, we use linear graph filters, which are extremely sparse. These design choices make our approach extremely efficient. To empirically validate the training efficiency of our method, we report the training time and storage cost of our methods and comparison methods. The results are presented in Table~\ref{tab:train_time} and Table~\ref{tab:train_gpu}. The method \textit{Student} denotes that we train the student model without knowledge distillation.

From the results, we find that:
\begin{itemize}[leftmargin=*]
    \item Although all knowledge distillation methods inevitably increase training costs, feature-based knowledge distillation methods bring about the least time cost. We believe this is because, in response-based distillation methods for recommender systems, it is often necessary to non-uniformly sample from all items, greatly increasing training time. For relation-based distillation methods, computing and storing huge similarity matrices and using them to compute distillation losses lead to an increase in both training time and storage cost.
    \item By comparing feature-based distillation methods (i.e., DE, FitNet, and \ourmethod), we find that DE has a higher training time and storage cost than the other two methods. We attribute this to introducing multiple projectors and a selection network in DE, all of which increase the storage cost and training time. In contrast, our approach adds only an extremely sparse graph filter. Thanks to the well-established tools for dealing with sparse matrices in various machine learning libraries, the storage and computation of this filter have a very low cost.
\end{itemize}

\subsection{Inference Efficiency}\label{sec:inference_eff}

In this section, we use our method to train students of different sizes. Then, we report their inference efficiency and recommendation performance. All results are obtained by testing with PyTorch on GeForce RTX 3090 GPU.

From the results in Table~\ref{tab:effi}, it is obvious that our method can benefit students of all sizes. Moreover, we achieve comparable performance to the teacher with only 10\% of the number of parameters and about a quarter of the inference latency. It demonstrates that by distilling with our method, we can significantly improve the inference efficiency of existing recommendation models. When the student is half the size of the teacher, the student can significantly outperform the teacher with only half the storage cost and greatly increase the inference efficiency. Moreover, with the same number of parameters, our method significantly outperforms the teacher, demonstrating that \ourmethod is also effective in improving existing recommendation models.

\subsection{Different Choices of $h(\lambda)$}\label{sec:h_lambda}

\begin{table}[htbp]
    \caption{Performance comparison for alternative design choices of $h(\lambda)$.}
    \label{tab:h_lambda}
    \centering
    \begin{tabular}{c|c|cc|cc} \toprule
        \multirow{2}{*}{Backbone} & \multirow{2}{*}{$h(\lambda)$} & \multicolumn{2}{c|}{Gowalla} & \multicolumn{2}{c}{Yelp}\\
         & & R@20 & N@20 & R@20 & N@20\\
        \midrule
        \multirow{2}{*}{BPRMF} & linear & 0.1494 & 0.1039 & 0.0669 & 0.0344\\
         & quadratic & 0.1502 & 0.1044 & 0.0673 & 0.0343\\
        \midrule
        \midrule
        \multirow{2}{*}{LightGCN} & linear & 0.1712 & 0.1175 & 0.0702 & 0.0359\\
         & quadratic & 0.1721 & 0.1179 & 0.0703 & 0.0355\\
        \midrule
        \midrule
        \multirow{2}{*}{SimpleX} & linear & 0.1739 & 0.1110 & 0.0758 & 0.0392\\
         & quadratic & 0.1746 & 0.1113 & 0.0762 & 0.0393\\
        \bottomrule
    \end{tabular}
\end{table}

In Eq.(\ref{eq:linear}), we set $h(\lambda)$ as a linear function for simplicity and efficiency. However, it is possible to design it into other higher-order forms. In this section, we verify empirically that the linear form, despite its simplicity, has a good performance. In Table~\ref{tab:h_lambda}, we give the results when $h(\lambda)$ is designed as a quadratic function, i.e., $h(\lambda)=a\lambda^2+b\lambda+1$. Note that we similarly restrict its shape to make it decrease in the domain of definition. From the results, we find that modifying $h(\lambda)$ to a quadratic function can slightly improve recommendation accuracy on Gowalla, while on Yelp it performs similarly to when it is a linear function. Therefore, we believe that the linear function of $h(\lambda)$ can be used as a strong option.

\section{Limitations}\label{sec:limit}
Although our method is efficient and effective in highlighting important knowledge, we have used only linear and quadratic filters. We did this design to keep the graph filter as sparse as possible to reduce the complexity of our method. However, this may result in a filter whose performance is not optimal. Therefore, how to design more complex filters to retain important knowledge better while removing unimportant knowledge would be an interesting research direction.

\section{Conclusion}
This paper analyses the feature-based knowledge distillation for recommendation from the frequency perspective. By defining different types of knowledge as different frequency components of features, we show that regular feature-based knowledge distillation equally minimizes losses on all types of knowledge. We then analyze how this equal loss weight allocation method overlooks important knowledge. Based on this, we propose emphasizing the important knowledge by redistributing knowledge weights and proposing a lightweight knowledge reweighting method. We theoretically and empirically validate the effectiveness of our method.

\section{Acknowledgments}
This work was supported in part by the National Natural Science Foundation of China under Grant 92270119, Grant 62072182, and the Key Laboratory of Advanced Theory and Application in Statistics and Data Science, Ministry of Education.

\bibliographystyle{ACM-Reference-Format}
\balance
\bibliography{refs}

\appendix

\section{Appendix}

\subsection{Prospects for applications on Relation-based KD}\label{sec:prospect}
In this paper, we focus on feature-based knowledge distillation methods. However, since relation-based knowledge distillation methods (e.g., HTD~\cite{kang2021topology}) are also closely related to features, our theoretical analysis can also be applied to them.
\begin{theorem}~\label{theorem-relation-based}
    Given the widely used relation-based distillation loss
    \begin{align}
        \mathcal{L}_{RD}=\|\mathbf{SS}^\top-\mathbf{TT}^\top\|_F^2
    \end{align}
    It can be decomposed into losses on each pair of frequency bands.
    Formally, we have
    \begin{align*}
        \|\mathbf{SS}^\top-\mathbf{TT}^\top\|_F^2=\sum_k\sum_p&\|\left(\mathbf{u}_k\mathbf{u}_k^\top\mathbf{S}\right)\left(\mathbf{u}_p\mathbf{u}_p^\top\mathbf{S}\right)^\top\\
        &\quad-\left(\mathbf{u}_k\mathbf{u}_k^\top\mathbf{T}\right)\left(\mathbf{u}_p\mathbf{u}_p^\top\mathbf{T}\right)^\top\|_F^2.
    \end{align*}
\end{theorem}
The proof is provided in Appendix~\ref{app:theorem3}. According to Theorem~\ref{theorem-relation-based}, the relation-based distillation loss can also be decomposed onto different frequency bands, which allows us to analyze relation-based methods using the ideas in our paper. Moreover, our experiments show that the relation-based approach (i.e., HTD~\cite{kang2021topology}) performs similarly to feature-based methods, and both are inferior to response-based methods. Therefore, we believe that extending the analytical approach of this paper to relation-based methods would be a meaningful research direction.

\subsection{Proof of Theorem~\ref{theorem-freq-decomp}}\label{app:theorem1}
In Theorem~\ref{theorem-freq-decomp}, we claim that the feature-based distillation loss $\mathcal{L}_{FD}=\|Proj(\mathbf{S})-\mathbf{T}\|_F^2$ can be decomposed into losses on different frequency bands. In this section, we will provide the proof.
\begin{proof}
\begin{align*}
    \mathcal{L}_{FD}=&\|Proj(\mathbf{S})-\mathbf{T}\|_F^2\\
    =&\|\mathbf{U}\mathbf{I}\mathbf{U}^\top (Proj(\mathbf{S})-\mathbf{T})\|_F^2\\
    =&\left\|\sum_{k=1}^{|\mathcal{V}|}\underbrace{\mathbf{u}_k\mathbf{u}_k^\top(Proj(\mathbf{S})-\mathbf{T})}_{\mathbf{Q}_k}\right\|_F^2\\
    =&tr\left(\left(\sum_{k=1}^{|\mathcal{V}|}\mathbf{Q}_k\right)^\top\left(\sum_{k=1}^{|\mathcal{V}|}\mathbf{Q}_k\right)\right)\\
    =&tr\left(\sum_{k=1}^{|\mathcal{V}|}{\mathbf{Q}_k}^\top \mathbf{Q}_k\right)+\sum_{i\neq j}tr\left({\mathbf{Q}_i}^\top \mathbf{Q}_j\right)\\
    =&\sum_{k=1}^{|\mathcal{V}|}\|\mathbf{Q}_k\|_F^2+\sum_{i\neq j}tr\left((Proj(\mathbf{S})-\mathbf{T})^\top \mathbf{u}_i\mathbf{u}_i^\top \mathbf{u}_j\mathbf{u}_j^\top (Proj(\mathbf{S})-\mathbf{T})\right)
\end{align*}
Since $\forall i\neq j,\mathbf{u}_i^\top \mathbf{u}_j=\mathbf{0}$, we have 
\begin{align*}
   \mathcal{L}_{FD}=&\sum_{k=1}^{|\mathcal{V}|}\|\mathbf{Q}_k\|_F^2\\
    =&\sum_{k=1}^{|\mathcal{V}|}\|\mathbf{u}_k\mathbf{u}_k^\top(Proj(\mathbf{S})-\mathbf{T})\|_F^2\\
    =&\sum_{k=1}^{|\mathcal{V}|}\|Proj(\tilde{\mathbf{S}}_k)-\tilde{\mathbf{T}}_k\|_F^2
\end{align*}
\end{proof}

\subsection{Proof of Theorem~\ref{theorem-filter}}\label{app:theorem2}
In Theorem~\ref{theorem-filter}, we assert that computing distillation loss using filtered features is equivalent to reweighting the distillation loss on each frequency band. In this section, we will provide the proof.
\begin{proof}
Given a graph filter $\mathcal{H}(\tilde{\mathbf{L}})$ defined as follows:
    \begin{align*}
        \mathcal{H}(\tilde{\mathbf{L}})=\mathbf{U}\text{Diag}(h(\lambda_1), h(\lambda_2), \cdots, h(\lambda_{|\mathcal{V}|})\mathbf{U}^\top.
    \end{align*}
We have
\begin{align*}
    \mathcal{L}_{FreqD}&=\|\mathcal{H}(\tilde{\mathbf{L}})Proj(\mathbf{S})-\mathcal{H}(\tilde{\mathbf{L}})\mathbf{T}\|_F^2\\
    &=\left\|\sum_{i=1}^{|\mathcal{V}|}h(\lambda_i)\mathbf{u}_i\mathbf{u}_i^\top(Proj(\mathbf{S})-\mathbf{T})\right\|_F^2\\
    &=tr(\left(\sum_{i=1}^{|\mathcal{V}|}h(\lambda_i)\mathbf{u}_i\mathbf{u}_i^\top(Proj(\mathbf{S})-\mathbf{T})\right)^\top\\
    &\quad\quad\cdot\left(\sum_{i=1}^{|\mathcal{V}|}h(\lambda_i)\mathbf{u}_i\mathbf{u}_i^\top(Proj(\mathbf{S})-\mathbf{T})\right))\\
    &=tr\left(\sum_{i}h_i^2(\lambda)\left(\mathbf{u}_i\mathbf{u}_i^\top(Proj(\mathbf{S})-\mathbf{T})\right)^\top\left(\mathbf{u}_i\mathbf{u}_i^\top(Proj(\mathbf{S})-\mathbf{T})\right)\right)\\
    &+tr\left(\sum_{i\neq j}h(\lambda_i)h(\lambda_j)(Proj(\mathbf{S})-\mathbf{T})^\top\mathbf{u}_i\mathbf{u}_i^\top\mathbf{u}_j\mathbf{u}_j^\top(Proj(\mathbf{S})-\mathbf{T})\right)\\
    &=tr\left(\sum_{i}h_i^2(\lambda)\left(\mathbf{u}_i\mathbf{u}_i^\top(Proj(\mathbf{S})-\mathbf{T})\right)^\top\left(\mathbf{u}_i\mathbf{u}_i^\top(Proj(\mathbf{S})-\mathbf{T})\right)\right)\\
    &=\sum_{i}h^2(\lambda_i)tr\left(\left(\mathbf{u}_i\mathbf{u}_i^\top(Proj(\mathbf{S})-\mathbf{T})\right)^\top\left(\mathbf{u}_i\mathbf{u}_i^\top(Proj(\mathbf{S})-\mathbf{T})\right)\right)\\
    &=\sum_{i=1}^{|\mathcal{V}|}h^2(\lambda_i)\|\mathbf{u}_i\mathbf{u}_i^\top(Proj(\mathbf{S})-\mathbf{T})\|_F^2\\
    &=\sum_{i=1}^{|\mathcal{V}|}h^2(\lambda_i)\|(Proj(\tilde{\mathbf{S}}_i)-\tilde{\mathbf{T}}_i)\|_F^2
\end{align*}
\end{proof}

\subsection{Proof of Theorem~\ref{theorem-relation-based}}\label{app:theorem3}
In Theorem~\ref{theorem-relation-based}, we claim that the loss used in some relation-based knowledge distillation methods can also be decomposed into losses on different frequency bands. More specifically, the $\mathcal{L}_{RD}=\|\mathbf{SS}^\top-\mathbf{TT}^\top\|_F^2$ consists of distillation loss on each pair of frequency. In this section, we will provide the proof.
\begin{proof}
\begin{align}
    \mathcal{L}_{RD}&=\|\mathbf{SS}^\top-\mathbf{TT}^\top\|_F^2\\
    &=tr\left((\mathbf{SS}^\top-\mathbf{TT}^\top)^\top(\mathbf{SS}^\top-\mathbf{TT}^\top)\right)\\
    &=tr(\mathbf{SS}^\top\mathbf{SS}^\top) + tr(\mathbf{TT}^\top\mathbf{TT}^\top)-tr(\mathbf{TT}^\top\mathbf{SS}^\top)-tr(\mathbf{SS}^\top\mathbf{TT}^\top)\label{eq:4tr}
\end{align}
Since the derivation of these four terms above is almost identical, we will focus on the first term only.
\begin{align*}
    tr(\mathbf{SS}^\top\mathbf{SS}^\top)&=tr\left(\mathbf{SS}^\top\mathbf{SS}\mathbf{UIU}^\top)\right)\\
    &=tr\left(\mathbf{SS}^\top\mathbf{SS}\left(\sum_{k}\mathbf{u}_k\mathbf{u}_k^\top\right)\right)\\
    &=\sum_k tr\left(\mathbf{SS}^\top\mathbf{SS}\mathbf{u}_k\mathbf{u}_k^\top\right)\\
    &=\sum_k tr\left(\mathbf{SS}^\top\mathbf{SS}\left(\mathbf{u}_k\mathbf{u}_k^\top\right)\left(\mathbf{u}_k\mathbf{u}_k^\top\right)\right)\\
    &=\sum_k tr\left(\left(\mathbf{u}_k\mathbf{u}_k^\top\right)\mathbf{SS}^\top\mathbf{SS}\left(\mathbf{u}_k\mathbf{u}_k^\top\right)\right)\\
    &=\sum_k tr\left(\left(\mathbf{u}_k\mathbf{u}_k^\top\right)\mathbf{SS}^\top\left(\sum_p\mathbf{u}_p\mathbf{u}_p^\top\right)\mathbf{SS}\left(\mathbf{u}_k\mathbf{u}_k^\top\right)\right)\\
    &=\sum_k\sum_p tr\left(\left(\mathbf{u}_k\mathbf{u}_k^\top\right)\mathbf{SS}^\top\left(\mathbf{u}_p\mathbf{u}_p^\top\right)\left(\mathbf{u}_p\mathbf{u}_p^\top\right)\mathbf{SS}\left(\mathbf{u}_k\mathbf{u}_k^\top\right)\right)\\
\end{align*}
For simplicity, let's define $\mathbf{S}_k:=\mathbf{u}_k\mathbf{u}_k^\top\mathbf{S}$ and $\mathbf{S}_p:=\mathbf{u}_p\mathbf{u}_p^\top\mathbf{S}$.

Then, we have
\begin{align*}
    tr(\mathbf{SS}^\top\mathbf{SS}^\top)&=\sum_k\sum_p tr(\mathbf{S}_k\mathbf{S}_p^\top\mathbf{S}_p\mathbf{S}_k^\top)
\end{align*}
Similarly, we can derive that
\begin{align*}
    tr(\mathbf{TT}^\top\mathbf{TT}^\top)&=\sum_k\sum_ptr(\mathbf{T}_k\mathbf{T}_p^\top\mathbf{T}_p\mathbf{T}_k^\top)\\
    tr(\mathbf{TT}^\top\mathbf{SS}^\top)&=\sum_k\sum_ptr(\mathbf{T}_k\mathbf{T}_p^\top\mathbf{S}_p\mathbf{S}_k^\top)\\
    tr(\mathbf{SS}^\top\mathbf{TT}^\top)&=\sum_k\sum_ptr(\mathbf{S}_k\mathbf{S}_p^\top\mathbf{T}_p\mathbf{T}_k^\top)
\end{align*}

By substituting the above results back into Eq.(\ref{eq:4tr}), we have
\begin{align*}
    \mathcal{L}_{RD}&=\sum_k\sum_p tr\left(\mathbf{S}_k\mathbf{S}_p^\top\mathbf{S}_p\mathbf{S}_k^\top+\mathbf{T}_k\mathbf{T}_p^\top\mathbf{T}_p\mathbf{T}_k^\top-\mathbf{T}_k\mathbf{T}_p^\top\mathbf{S}_p\mathbf{S}_k^\top-\mathbf{S}_k\mathbf{S}_p^\top\mathbf{T}_p\mathbf{T}_k^\top\right)\\
    &=\sum_k\sum_ptr\left(\left(\mathbf{S}_k\mathbf{S}_p^\top-\mathbf{T}_k\mathbf{T}_p^\top\right)\left(\mathbf{S}_k\mathbf{S}_p^\top-\mathbf{T}_k\mathbf{T}_p^\top\right)^\top\right)\\
    &=\sum_k\sum_p\|\mathbf{S}_k\mathbf{S}_p^\top-\mathbf{T}_k\mathbf{T}_p^\top\|_F^2\\
    &=\sum_k\sum_p\|\left(\mathbf{u}_k\mathbf{u}_k^\top\mathbf{S}\right)\left(\mathbf{u}_p\mathbf{u}_p^\top\mathbf{S}\right)^\top-\left(\mathbf{u}_k\mathbf{u}_k^\top\mathbf{T}\right)\left(\mathbf{u}_p\mathbf{u}_p^\top\mathbf{T}\right)^\top\|_F^2\\
\end{align*}
\end{proof}

\subsection{Hyperparameter Analysis}\label{sec:hyper}

\begin{figure}[htbp]
  \centering
  \includegraphics[width=1.0\linewidth]{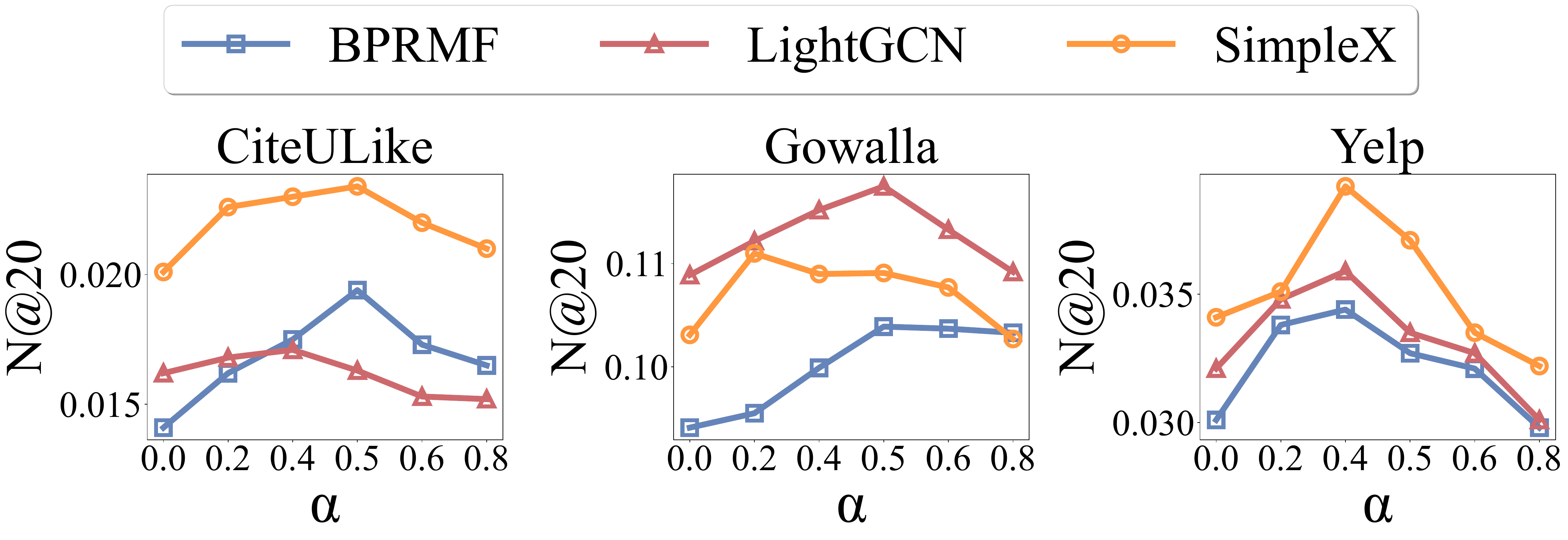}
  \caption{Effects of $\alpha$.}
  \label{fig:hyper_alpha}
\end{figure}

\begin{figure}[htbp]
  \centering
  \includegraphics[width=1.0\linewidth]{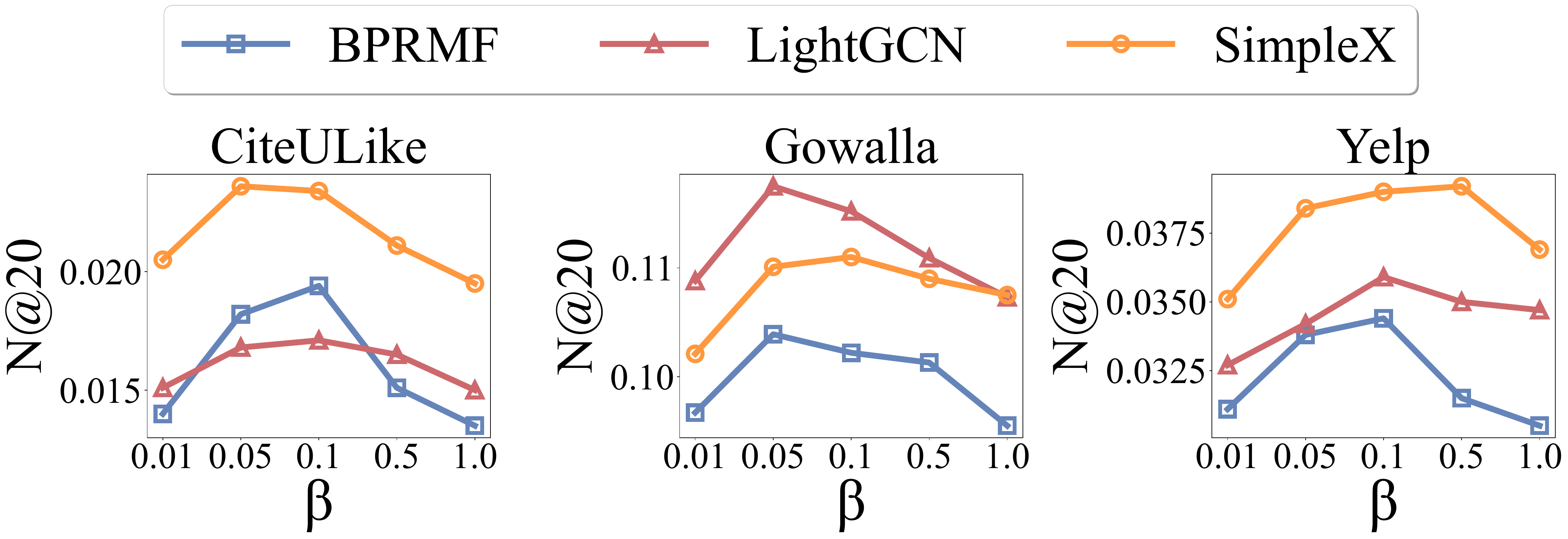}
  \caption{Effects of $\beta$.}
  \label{fig:hyper_beta}
\end{figure}

This section analyses the effects of hyperparameters of \ourmethod in terms of NDCG@20.

\textbf{Effects of $\alpha$.}
In Eq.(\ref{eq:linear}), we define the graph filter in \ourmethod as a linear function of the graph Laplacian matrix and introduce hyperparameter $\alpha$. The effects of $\alpha$ is reported in Figure~\ref{fig:hyper_alpha}. The experimental results illustrate that too large or too small $\alpha$ results in performance dropping. This is because too large an $\alpha$ leads to too much loss of high-frequency knowledge, while too small an $\alpha$ can not motivate the student model to focus on low-frequency knowledge. The optimal value of $\alpha$ is usually between $0.4$ and $0.5$.

\textbf{Effects of $\beta$.}
In Eq.(\ref{eq:final_loss}), hyperparameter $\beta$ represents the weight of the feature-based distillation loss. The effects of $\beta$ is shown in Figure~\ref{fig:hyper_beta}. Because the types of loss functions of the proposed methods differ from those of the base models, it is important to balance the losses by properly setting $\beta$. In our experiments, we find that setting $\beta$ to $0.1$ usually resulted in good performance.

\end{document}